\documentclass{jaa}
\usepackage{natbib}
\usepackage{amsmath}
\usepackage{amssymb}
\usepackage{graphicx}
\usepackage{booktabs}
\usepackage{hyperref}
\usepackage{float}

\begin{document}\sloppy

\title{Testing the Generalization and Domain Stability of Compact Feature Representations for Photometric Supernova Classification}

\author{Anurag Garg\textsuperscript{1, *}}
\affilOne{\textsuperscript{1}Ministry of Education, Abu Dhabi, UAE.\\}

\twocolumn[{

\maketitle

\corres{anurag.garg@moe.sch.ae}

\msinfo{20 June 2026}{}{}

\begin{abstract}
Photometric classification of supernovae increasingly requires models that are not only accurate within a single survey, but also stable under changes in cadence, noise properties, filter coverage, and survey domain. In this work, we test the generalization and domain stability of a compact 16-feature representation for Type Ia supernova classification. The feature set consists of physically interpretable light-curve descriptors measuring brightness scale, color structure, variability, and temporal evolution. Using the Supernova Photometric Classification Challenge (SPCC) dataset as the reference domain, we first confirm that the compact XGBoost classifier preserves strong within-survey performance, achieving an F1 score of approximately 0.844 and a PR-AUC of approximately 0.928 on the held-out test set. We then evaluate robustness under alternate classifiers, repeated resampling, synthetic feature perturbations, missing-band proxies, shortened temporal coverage, and cross-survey transfer to PLAsTiCC.

The compact representation remains stable under resampling and moderate perturbations, but external transfer reveals a substantially larger limitation. Direct SPCC$\rightarrow$PLAsTiCC transfer produces a marked drop in performance, and class-conditional centroid analysis shows that the SPCC and PLAsTiCC Type Ia populations do not occupy the same region of the compact feature space. The cross-survey Ia centroid shift is larger than the Ia/non-Ia separation within either survey, indicating that the transfer gap is caused by feature-space domain shift rather than classifier instability alone. These results show that compact physically interpretable features are robust within a survey and useful for diagnostic analysis, but are not automatically survey-invariant. Cross-survey deployment requires feature-space harmonization, domain adaptation, or restriction to a smaller set of survey-stable features.
\end{abstract}


}]

\section{Introduction}
Photometric classification of supernovae has become essential for modern time-domain astronomy, where the number of detected transients exceeds the capacity for spectroscopic confirmation. Large surveys require automated methods that can identify Type~Ia supernovae from multi-band light curves while operating under irregular cadence, heterogeneous noise, incomplete filter coverage, and evolving survey conditions. The Supernova Photometric Classification Challenge (SPCC) provided an early standardized benchmark for this problem and motivated a wide range of machine-learning approaches for photometric supernova classification \cite{Kessler2010}.

Feature-based machine-learning methods have shown that physically motivated light-curve summaries can achieve strong classification performance. \cite{Lochner2016} compared several feature-extraction techniques and classifiers, demonstrating that carefully constructed features can perform competitively with more complex methods. \cite{Karpenka2013} used parametric light-curve fitting to derive compact descriptors for automated supernova classification. Such approaches are attractive because the input variables can often be connected directly to physical properties of the transient, including brightness scale, color evolution, variability, and temporal structure.

Deep-learning approaches have also been applied successfully to supernova light curves, including recurrent and convolutional architectures that operate directly on sequences or image-like representations \cite{Charnock2017,Moller2019,Pasquet2019}. These models can achieve strong performance, but their internal representations are generally harder to interpret and may be sensitive to survey-specific cadence, noise, and selection effects. This motivates continued investigation of compact feature representations that are both physically interpretable and computationally efficient.

A compact physically interpretable representation for photometric supernova classification was introduced by \cite{Garg2025}, demonstrating that a small set of physically motivated descriptors could preserve most of the predictive performance of substantially larger feature collections. Subsequent work by \cite{Garg2026} examined the internal structure of this representation through feature-ablation and subset analyses, showing that temporal, color, and variability information contribute unequally to classification performance and identifying a stable compact core.

However, strong performance within one survey does not by itself imply survey portability. The previous studies established that the compact representation is effective and interpretable within its native survey domain, but they did not address whether the same feature space remains stable when applied across surveys with different observing strategies, cadence patterns, filter responses, and noise characteristics.

The present work addresses that question directly. Rather than introducing a new feature representation, we use the established compact 16-feature framework as a reference model and evaluate its stability under model changes, resampling, controlled feature perturbations, missing-band proxies, shortened temporal coverage, and cross-survey transfer to PLAsTiCC. The goal is to determine whether the compact representation is merely accurate within SPCC or whether it remains stable under realistic domain shifts.

This framing allows both positive and negative outcomes to be scientifically informative. If the compact representation remains stable, it would support the case for survey-portable interpretable features. If it fails under external transfer, the failure identifies where interpretability and survey invariance diverge. In particular, class-conditional centroid analysis can test whether Type~Ia events from different surveys occupy the same region of the compact feature space, or whether transfer is limited before classifier optimization is even considered.

The paper is organized as follows. Section~2 describes the SPCC and PLAsTiCC datasets used in this work. Section~3 presents the compact feature representation and feature-construction procedure. Section~4 summarizes the classification model and evaluation metrics. Section~5 gives the frozen compact-baseline performance. Section~6 presents the robustness and domain-stability experiments, including cross-survey transfer and class-conditional centroid analysis. Section~7 discusses the implications for interpretable feature design and survey portability. Section~8 summarizes the conclusions.

\section{Dataset}
The primary dataset used in this work is the Supernova Photometric Classification Challenge (SPCC) dataset \cite{Kessler2010}. Raw observations are stored as DES light-curve files containing simulated supernova events with associated metadata, including object identifier, simulated class label, simulated redshift, observation time (MJD), photometric band, flux, and flux uncertainty.

The SPCC data are loaded and cleaned using the preprocessing workflow described in \cite{Garg2026}. Observations are sorted in time, non-finite values are removed, and events with invalid metadata are rejected before feature extraction.

For domain-generalization tests, the PLAsTiCC simulated survey dataset is also used. The PLAsTiCC data are read from the official training tables together with the associated metadata file, and the same compact feature schema is reconstructed directly from the raw observations in order to ensure consistency between datasets.

\begin{table}[H]
\centering
\caption{Datasets used in this study.}
\label{tab:dataset_compare}
\begin{tabular}{p{0.28\columnwidth}p{0.28\columnwidth}p{0.28\columnwidth}}
\hline
Property & SPCC & PLAsTiCC \\
\hline
Survey analogue & DES-like & LSST-like \\
Role in this study & Reference domain & External transfer domain \\
Bands used & $g,r,i,z$ & $g,r,i,z$ \\
Data source & Challenge simulations & PLAsTiCC simulations \\
Feature extraction & Native compact pipeline & Reconstructed compact pipeline \\
Purpose & Training and baseline evaluation & Domain-shift evaluation \\
\hline
\end{tabular}
\end{table}

All experiments use a stratified split protocol with a test fraction of 0.2, followed by a validation fraction of 0.2 applied to the remaining training data, giving an effective train/validation/test split of 64/16/20. This split protocol is fixed across all experiments to allow consistent comparison of results.

\section{Compact Feature Representation}

The compact feature representation used in this work follows the physically interpretable feature model introduced by \cite{Garg2026}.

\begin{table*}[t]
\centering
\caption{Compact 16-feature representation used throughout this work.}
\label{tab:compact_features}
\begin{tabular}{lll}
\hline
Category & Feature & Physical interpretation \\
\hline

Brightness &
$i$-band peak flux &
Approximate peak luminosity proxy \\

Brightness &
$i$-band mean flux &
Average brightness level \\

Brightness &
$i$-band flux standard deviation &
Brightness variability \\

Brightness &
Amplitude$_i$ &
Peak-to-trough brightness range \\

Temporal &
Duration &
Active transient lifetime \\

Temporal &
$t_{\rm peak,g}$ &
Time of maximum flux in $g$ band \\

Temporal &
$t_{\rm peak,r}$ &
Time of maximum flux in $r$ band \\

Temporal &
$t_{\rm peak,i}$ &
Time of maximum flux in $i$ band \\

Temporal &
$t_{\rm peak,z}$ &
Time of maximum flux in $z$ band \\

Brightness &
Mean flux$_g$ &
Representative brightness in $g$ band \\

Brightness &
Mean flux$_r$ &
Representative brightness in $r$ band \\

Brightness &
Mean flux$_i$ &
Representative brightness in $i$ band \\

Brightness &
Mean flux$_z$ &
Representative brightness in $z$ band \\

Color &
$C_{g-r}$ &
Blue-to-red color evolution \\

Color &
$C_{r-i}$ &
Intermediate color index \\

Color &
$C_{i-z}$ &
Red-band color index \\

\hline
\end{tabular}
\end{table*}

Only the $g$, $r$, $i$, and $z$ photometric bands are used. For each event, an active observation window is defined using measurements with positive flux, positive flux uncertainty, and signal‑to‑noise ratio greater than or equal to three. If no such measurements exist, the full observation set is used. The active window is therefore defined as the subset of observations satisfying $F>0$, $\sigma_F>0$, and $F/\sigma_F \ge 3$, where $F$ is the measured flux and $\sigma_F$ is the reported flux uncertainty.

Within the active window, per‑band peak flux, mean flux, standard deviation, and time of peak are computed. The amplitude in the $i$ band is defined as the difference between the maximum and minimum flux within the active window, and the event duration is defined as the time span of the active window.

Color features are constructed from representative positive flux values in each band. For each band, the mean of the strongest positive observations is used to define a stable flux estimate. Colors are then computed using magnitude‑style ratios between bands following the standard astronomical magnitude definition \cite{Ivezic2019}.

For two bands with representative flux values $F_a$ and $F_b$, the color feature is defined using a magnitude-style relation

\[
C_{a-b} = -2.5 \log_{10}\left(\frac{F_a}{F_b}\right)
\]

where the representative flux for each band is taken as the mean of the strongest positive observations within the active window.

The compact feature representation used here is identical to the one introduced in \cite{Garg2026}, where the full list of features, their physical interpretation, and the single‑feature ablation analysis are presented in detail. In the present work the feature definition itself is not modified; instead, the same compact representation is used to test stability under changes in data distribution, model choice, noise level, and training sample size. Flux‑like quantities are compressed using logarithmic scaling after non‑negative clipping, following common feature‑scaling practice in machine learning \cite{Hastie2009}. Signed mean‑flux quantities are transformed using a signed logarithmic mapping defined in this work in order to preserve the sign information while limiting the dynamic range, and color features are clipped to a finite interval to prevent extreme ratios caused by low‑signal observations from dominating the feature space.

Flux-like quantities are compressed using a logarithmic transform

\[
x' = \log_{10}(1 + x)
\]

after non-negative clipping. Signed quantities are transformed using

\[
x' = \mathrm{sign}(x)\,\log_{10}(1 + |x|)
\]

and color features are clipped to the interval $[-5, 5]$ in order to limit the influence of extreme ratios caused by low signal observations.

The same compact schema is rebuilt directly from raw SPCC and PLAsTiCC observations using the processing scripts in the public repository to ensure that the representation does not depend on precomputed features and can be reproduced from raw survey data.

In addition to the general preprocessing described above, the compact feature construction defined in \cite{Garg2026,GargRepo2026} requires that all four photometric bands $g$, $r$, $i$, and $z$ provide usable positive flux measurements for color computation. The representative flux in each band is defined from the strongest positive observations within the active window, and the color features
$C_{g-r}$, $C_{r-i}$, and $C_{i-z}$ require valid flux values in all four bands. An event is therefore rejected at the feature-construction stage if any band lacks sufficient positive flux support.

The event rejection rule operates at two levels in the processing pipeline. First, during raw-event cleaning, events are rejected if core metadata are missing, invalid, or if too few valid observations remain after numeric filtering. Second, after feature construction, events are rejected if the compact feature definition cannot be completed because one or more bands do not provide usable measurements for the color features.

This requirement significantly affects the final sample size when the same feature schema is applied to different surveys. In particular, enforcing the four-band consistency condition removes objects with incomplete or weak multiband coverage, leading to a smaller but internally consistent dataset. The reduction in the final compact-feature table therefore reflects the stricter feature-definition requirement rather than any change in the sampling procedure.

\section{Classification Model and Evaluation Metrics}

The classification experiments were performed using the compact feature representation described in the previous section. All models were trained on the same stratified data split and evaluated using a common metric framework implemented in the public repository \cite{GargRepo2026}, using the scripts on branch \texttt{phase2-tier4-domain-generalization}.

\subsection{Classification task}

The problem is formulated as a binary classification task, with label $1$ assigned to Type~Ia supernovae and label $0$ assigned to non-Ia events. All classifiers operate on the sixteen-dimensional compact feature vector defined in \cite{Garg2026}.

Before fitting, input features are standardized using the mean and standard deviation computed from the training set. The same transformation is applied to validation and test data using the training-set statistics.

\subsection{Train, validation, and test split}

All experiments use a stratified split with $20\%$ of the data reserved as a held-out test set. The remaining $80\%$ is further split into training and validation subsets, with $20\%$ of the remaining data used for validation. This produces an effective split of $64\%$ training, $16\%$ validation, and $20\%$ test, which is kept fixed across all model comparisons.

Model selection is performed using the validation set. After the best model and hyperparameters are chosen, the classifier is refit on the combined training and validation data and evaluated on the held-out test set.

\subsection{Candidate classifiers}

The compact feature representation was evaluated with four classifier families in order to test whether the observed performance was specific to one model architecture or reflected information present in the feature space itself. The candidate models were chosen to span different decision-boundary assumptions:

\begin{itemize}
\item XGBoost gradient-boosted decision trees, representing a nonlinear ensemble model with strong performance on tabular data,
\item Random Forest, representing a bagged tree ensemble with reduced sensitivity to individual tree structure,
\item Logistic Regression, representing a linear baseline,
\item Support Vector Machine with radial-basis-function kernel, representing a nonlinear margin-based classifier.
\end{itemize}

This comparison was not intended as an exhaustive hyperparameter search across all possible algorithms. Its purpose was to check whether the compact 16-feature representation carries a stable classification signal under both linear and nonlinear modelling assumptions.

These models were compared under the same split and metric framework using the model-comparison script on the current repository branch.  
Because the dataset is strongly imbalanced, model selection was based primarily on precision–recall area under the curve (PR-AUC) rather than accuracy or ROC-AUC.  
PR-AUC directly evaluates the ranking quality of the positive class and is therefore more appropriate when the scientific objective is reliable identification of Type~Ia events under class imbalance, as discussed in \cite{Garg2025}.  
Under this criterion, XGBoost achieved the best overall performance and was therefore selected as the primary classifier for all subsequent experiments.

\begin{table}[H]
\centering
\caption{Model comparison on the compact feature set.}
\label{tab:model_compare}
\begin{tabular}{lccc}
\hline
Model & F1 & ROC-AUC & PR-AUC \\
\hline
XGBoost & 0.844 & 0.977 & 0.928 \\
Random Forest & 0.824 & 0.974 & 0.924 \\
Logistic Regression & 0.714 & 0.895 & 0.608 \\
SVM (RBF) & 0.835 & 0.970 & 0.886 \\
\hline
\end{tabular}
\end{table}

The numerical results are summarized in Table~\ref{tab:model_compare}.  
These values are taken directly from the model-comparison output of the repository \cite{GargRepo2026}.

\subsection{XGBoost configuration}

The final classifier used for the compact-feature workflow is XGBoost with the following configuration:

\begin{itemize}
\item Objective: \texttt{binary:logistic}
\item Evaluation loss: logloss
\item Tree method: \texttt{hist}
\item Maximum boosting rounds: 400
\item Early stopping: 30 rounds
\item Class imbalance handling:
\[
\texttt{scale\_pos\_weight} = \frac{N_{\text{negative}}}{N_{\text{positive}}}
\]
\end{itemize}

The hyperparameter grid tested was

\begin{itemize}
\item max\_depth = 3, eta = 0.05, subsample = 0.8, colsample\_bytree = 0.8,
      min\_child\_weight = 1.0, lambda = 1.0
\item max\_depth = 4, eta = 0.05, subsample = 0.9, colsample\_bytree = 0.9,
      min\_child\_weight = 1.0, lambda = 1.0
\item max\_depth = 5, eta = 0.03, subsample = 0.8, colsample\_bytree = 0.8,
      min\_child\_weight = 2.0, lambda = 1.5
\end{itemize}

The best configuration was selected using validation PR-AUC.

\subsection{Evaluation metrics}

The following metrics are computed for all experiments:

\begin{itemize}
\item Accuracy
\item Precision
\item Recall
\item F1 score
\item ROC-AUC
\item PR-AUC
\end{itemize}

Predicted probabilities are obtained from the classifier.  
Precision, recall, and F1 are computed using a threshold of $0.5$ on the predicted probability.

The F1 score is defined as

\[
F1 = \frac{2 \, \text{precision} \times \text{recall}}
           {\text{precision} + \text{recall}}
\]

ROC-AUC is computed from the ranked prediction scores, and PR-AUC is computed as the average precision obtained from the precision–recall curve.

Model selection is based on PR-AUC rather than accuracy, because the dataset is class-imbalanced and correct ranking of Type~Ia candidates is more important than overall classification rate.

\section{Baseline Compact Model Performance}

The frozen reference model used throughout the later experiments is the compact 16-feature XGBoost baseline introduced in \cite{Garg2026}. That study established that the compact representation preserves essentially the full discriminative performance of the larger feature sets while improving interpretability and reducing feature-space complexity. In the present work, the same compact feature definition, split protocol, and evaluation workflow are retained in order to provide a stable reference point for all subsequent robustness tests.

Using the fixed stratified split described in Section~4, the final XGBoost compact baseline is selected on the validation set using PR-AUC, retrained on the combined training and validation data, and evaluated on the held-out test set. The resulting test-set metrics are summarized in Table~\ref{tab:compact_baseline_metrics}.

\begin{table}[H]
\centering
\caption{Frozen compact-baseline performance on the held-out test set.}
\label{tab:compact_baseline_metrics}
\begin{tabular}{lc}
\hline
Metric & Value \\
\hline
Accuracy & 0.916745 \\
Precision & 0.762887 \\
Recall & 0.944990 \\
F1 & 0.844230 \\
ROC-AUC & 0.976588 \\
PR-AUC & 0.927761 \\
\hline
\end{tabular}
\end{table}

The headline compact-model results, F1 $\approx 0.844$ and PR-AUC $\approx 0.928$, are consistent with the values reported in \cite{Garg2026}. The inclusion of the full metric spectrum here makes explicit that the compact model performs reasonably well across both threshold-dependent and threshold-independent summaries, rather than appearing strong only under a single preferred metric.

The compact baseline was also compared against larger feature representations using the same split, model, and evaluation workflow. The comparison is summarized in Table~\ref{tab:feature_set_compare}.

\begin{table}[H]
\centering
\caption{Performance comparison for different feature-set sizes.}
\label{tab:feature_set_compare}
\begin{tabular}{lccc}
\hline
Feature set & F1 & ROC-AUC & PR-AUC \\
\hline
31-feature full set & 0.8367 & 0.9764 & 0.9281 \\
30-feature working set & 0.8405 & 0.9764 & 0.9282 \\
16-feature compact set & 0.8442 & 0.9766 & 0.9278 \\
\hline
\end{tabular}
\end{table}

Relative to the 31-feature baseline, the compact model slightly improves the F1-score while leaving ROC-AUC essentially unchanged and reducing PR-AUC only marginally. This confirms that the compact representation preserves the discriminative performance of the larger feature sets while using a smaller and physically interpretable feature space.

These baseline results justify the use of the compact XGBoost model as the fixed reference for the subsequent analyses. Because the compact model is both physically interpretable and near-parity with the larger feature sets, later changes in performance can be attributed more directly to perturbations in model family, data quality, or survey domain rather than to instability in the baseline feature definition itself.



\subsection{Transferability test design}

The main checks used to support this transferability assessment are summarized below. The purpose is not to repeat the full numerical analysis, but to make clear that the final domain-shift conclusion is based on a sequence of tests rather than on a single failed external-transfer experiment.

These checks include both the experiments reported numerically in the main text and additional diagnostic trials used to rule out simpler explanations for the transfer gap.

The transferability assessment combined classifier-family comparison, resampling stability, perturbation tests, domain-swap transfer, mixed-domain training, feature-definition audits, normalization trials, transient-window harmonization, and class-conditional centroid analysis. A compact summary of these diagnostics is provided in Appendix~\ref{app:diagnostics}.

This staged design makes the final conclusion more conservative. The negative transfer result is not interpreted as a failure of the classifier alone, because the compact representation is stable under several internal tests. Instead, the combined experiments identify the remaining limitation as feature-space domain shift between surveys.


\section{Robustness Experiments}

Robustness was evaluated at three complementary levels: structural robustness of the compact representation, training robustness under alternate model and split choices, and domain robustness under controlled distribution shifts and external-survey transfer. The goal of this section is to determine whether the predictive power of the compact feature set is concentrated in a fragile subset of variables, tied to one classifier family or one favorable split, or instead remains stable under perturbations that mimic realistic survey limitations.

\subsection{Structural robustness of the compact representation}

The structural robustness of the compact representation has already been examined in detail in \cite{Garg2026}, where single-feature ablation, block-level ablation, subset-growth experiments, and reduced-core tests were presented for the same 16-feature compact model. That analysis showed that the classification signal is not carried by one fragile variable, that temporal information provides the strongest individual contribution, and that a reduced core of approximately ten features retains most of the full compact-model performance. The corresponding paper reports the single-feature ablation results on page~5, the block-level and subset-growth analysis on pages~5--6, and the minimal-core results on page~6. 

Because those structural tests are already the central contribution of \cite{Garg2026}, they are not repeated in full here. Instead, the present work takes the compact representation as a fixed input and focuses on a different question: whether that already-established compact feature set remains stable under changes in classifier family, resampling protocol, feature-space perturbation, and survey domain. In this sense, the structural robustness results from \cite{Garg2026} serve as the foundation for the broader robustness analysis reported below.

\subsection{Robustness to model choice, resampling, and feature-space perturbation}

The second set of experiments tests whether the compact representation depends on one particular classifier family, one fortunate train/test split, or unrealistically clean feature values. The model-comparison results in Table~\ref{tab:model_compare} already show that XGBoost is the strongest classifier on the compact feature set, but the relative performance of Random Forest and SVM remains close enough to indicate that the representation is not exclusively tied to one model class. Logistic Regression performs substantially worse, suggesting that the relevant decision boundary is not well described by a purely linear separator.

Stability to alternate resampling protocols was assessed through repeated $k$-fold cross-validation and repeated random stratified splits. The results are summarized in Table~\ref{tab:resampling_stability}. The mean F1-score remains close to the frozen baseline under both protocols, and the observed standard deviations are small, indicating that the compact representation is not strongly split-fragile.

\begin{table}[H]
\centering
\caption{Resampling stability of the compact XGBoost baseline.}
\label{tab:resampling_stability}
\begin{tabular}{lcccc}
\hline
Protocol & Runs & Mean F1 & F1 std. & PR-AUC \\
\hline
$k$-fold CV & 5 & 0.841 & 0.006 & 0.923 \\
Random split & 5 & 0.845 & 0.003 & 0.923 \\
\hline
\end{tabular}
\end{table}

The compact baseline was then stress-tested using feature-space perturbations that mimic reduced coverage, added noise, missing-band proxies, and shortened time sampling. The perturbation results are summarized in Table~\ref{tab:perturbation_robustness}. Moderate degradation is observed for reduced observations and for additive flux noise at the $+0.25\sigma$ level, while stronger degradation appears for larger noise, missing $z$-band proxies, and shortened time coverage. In particular, removing $z$-band proxies produces the largest F1 drop among these perturbations ($\Delta$F1 = -0.239891), showing that red-band information is especially important for stable classification.

\begin{table}[H]
\centering
\caption{Feature-space perturbation tests for the compact XGBoost baseline.}
\label{tab:perturbation_robustness}
\begin{tabular}{lcc}
\hline
Scenario & F1 & $\Delta$F1 \\
\hline
No perturbation & 0.844230 & +0.000000 \\
Reduced observations & 0.730242 & -0.113988 \\
Flux noise (+0.25$\sigma$) & 0.782128 & -0.062102 \\
Flux noise (+0.50$\sigma$) & 0.625885 & -0.218345 \\
Remove $z$-band proxies & 0.604339 & -0.239891 \\
Shortened time coverage & 0.704953 & -0.139277 \\
\hline
\end{tabular}
\end{table}

The reduced-core subsets were also retested under alternate seeds, models, and perturbations. Their mean F1-scores were 0.5929 for the top-5 subset, 0.6673 for the top-8 subset, and 0.7151 for the top-10 subset. Among the reduced representations, the 10-feature core is therefore the most robust compact subset beyond the original frozen split.

\subsection{Domain robustness and cross-survey transfer}

The strongest robustness test in the repository evaluates whether the compact representation survives controlled distribution shift and external-survey transfer. In these experiments, the frozen SPCC compact baseline is evaluated on SPCC-derived shifted domains and on compact features reconstructed from the PLAsTiCC survey data. The tested domains include additive noise, reduced temporal span, global flux scaling, removal of $z$-band information, removal of $i$-band information, and transfer to PLAsTiCC. The domain-swap results are summarized in Table~\ref{tab:domain_swap}.

\begin{table}[H]
\centering
\caption{Domain-swap performance of the frozen compact XGBoost baseline. Drops are measured relative to the SPCC test domain.}
\label{tab:domain_swap}
\begin{tabular}{lcc}
\hline
Test domain & F1 & Drop in F1 \\
\hline
SPCC & 0.8407 & 0.0000 \\
Noise & 0.8263 & -0.0179 \\
Short span & 0.8194 & -0.0248 \\
Flux scale & 0.7730 & -0.0713 \\
No $z$ band & 0.6921 & -0.1521 \\
No $i$ band & 0.6522 & -0.1920 \\
PLAsTiCC & 0.5168 & -0.3274 \\
\hline
\end{tabular}
\end{table}

\begin{figure}[H]
\centering
\caption{Domain-swap performance of the compact SPCC-trained classifier. Mild SPCC-like perturbations cause modest degradation, while missing-band conditions and PLAsTiCC transfer produce substantially larger drops.}
\includegraphics[width=0.95\columnwidth]{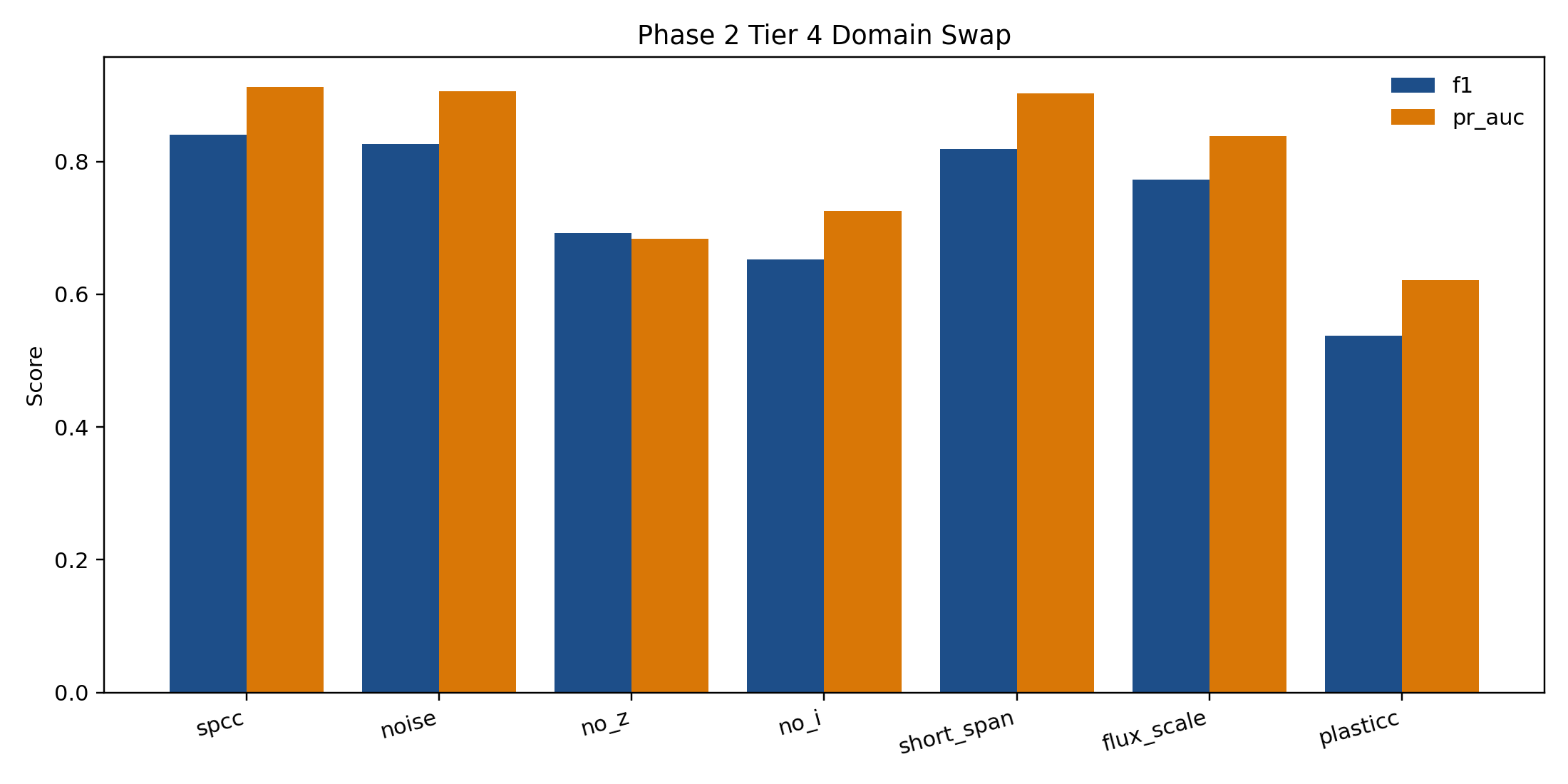}
\label{fig:domain_swap}
\end{figure}

These results, shown graphically in Fig.~\ref{fig:domain_swap}, show a graded degradation pattern rather than immediate collapse. Mild perturbations such as noise and shortened coverage reduce performance only modestly, while missing-band conditions are substantially more damaging. The hardest transfer is to PLAsTiCC, where the F1-score drops by 0.327406 relative to SPCC. This indicates that cross-survey transfer is weak when the classifier is trained only on the SPCC domain, even though the compact representation remains partially predictive.

To test whether the transfer gap arises entirely from feature failure or partly from train-domain mismatch, mixed-domain training experiments were also performed. Training on SPCC alone yields a mean F1 of 0.7047 across the evaluated domains. Adding shifted SPCC variants provides only small changes, but adding PLAsTiCC to the training mixture raises the mean F1 to 0.7468. This does not imply full survey invariance, but it does show that a substantial part of the cross-domain degradation is due to mismatch between training and deployment domains rather than complete breakdown of the compact representation.

\begin{figure}[h]
\centering
\caption{Mean cross-domain performance under different training-domain mixtures. Including PLAsTiCC-like examples improves transfer performance, indicating that part of the degradation arises from train/deployment domain mismatch rather than complete failure of the compact representation.}
\includegraphics[width=0.95\columnwidth]{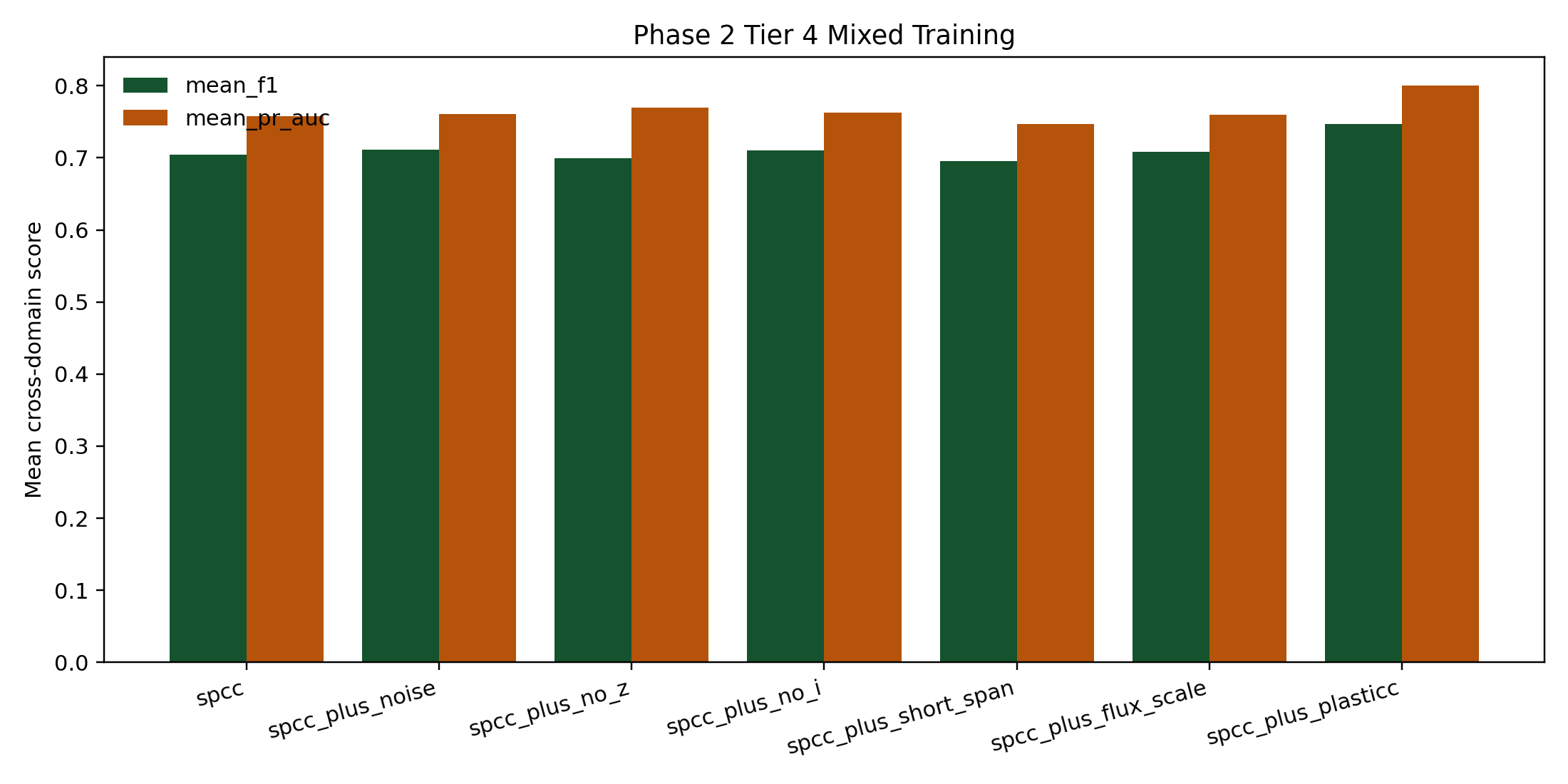}
\label{fig:mixed_training}
\end{figure}

Finally, reduced feature subsets were re-evaluated across domains. The mean F1-scores are 0.6751 for the top-5 subset, 0.7271 for the top-8 subset, 0.7791 for the top-10 subset, and 0.7983 for the full compact feature set. The full compact model therefore remains the most stable representation under domain shift, although the top-10 subset retains a substantial fraction of the domain-averaged performance.

Together, these results support three conclusions. 
\begin{enumerate}
    \item The compact signal is distributed across physically meaningful feature families rather than being carried by one fragile variable.
    \item The representation remains stable across resampling protocols and across multiple nonlinear classifier families.
    \item The compact model gradually degrades under controlled perturbations and moderate domain shift, while strong missing-band conditions and external-survey transfer remain the principal limitations of the current workflow.
\end{enumerate}

\subsection{Class-conditional centroid analysis}

To determine whether the remaining SPCC$\rightarrow$PLAsTiCC transfer limitation is caused primarily by classifier-boundary mismatch or by the feature representation itself, we performed a class-conditional centroid analysis in the standardized compact 16-feature space. The four centroids measured were SPCC Ia, PLAsTiCC Ia, SPCC non-Ia, and PLAsTiCC non-Ia.

The resulting centroid distances are shown in Table~\ref{tab:centroid_analysis}. Distances were computed after applying the same feature standardization used in the compact-model workflow.

\begin{figure}[h]
\centering
\caption{PCA projection of the four standardized class-conditional centroids. The separation between the SPCC and PLAsTiCC Type Ia centroids provides a visual diagnostic of the cross-survey feature-space shift.}
\includegraphics[width=0.95\columnwidth]{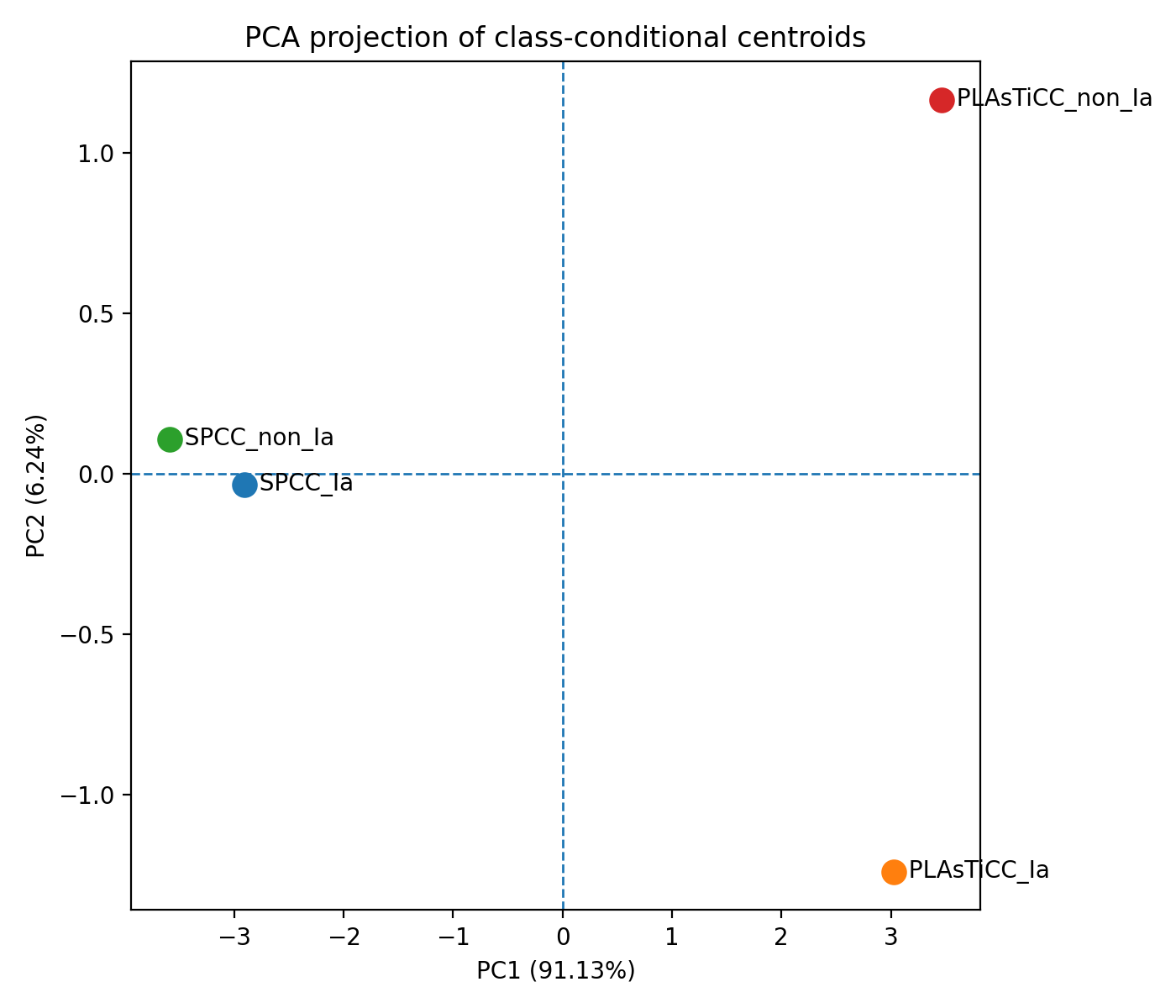}
\label{fig:centroid_pca}
\end{figure}

\begin{table}[H]
\centering
\caption{Class-conditional centroid distances in the standardized compact feature space.}
\label{tab:centroid_analysis}
\begin{tabular}{lc}
\hline
Quantity & Distance / ratio \\
\hline
SPCC Ia $\rightarrow$ PLAsTiCC Ia & 6.125 \\
SPCC non-Ia $\rightarrow$ PLAsTiCC non-Ia & 7.175 \\
SPCC Ia $\rightarrow$ SPCC non-Ia & 1.710 \\
PLAsTiCC Ia $\rightarrow$ PLAsTiCC non-Ia & 2.448 \\
Ia shift / SPCC class separation & 3.58$\times$ \\
Ia shift / PLAsTiCC class separation & 2.50$\times$ \\
\hline
\end{tabular}
\end{table}

\begin{figure}[H]
\centering
\caption{Feature-wise Ia centroid shift between PLAsTiCC and SPCC in the standardized compact feature space. Larger absolute shifts indicate features whose numerical distributions are less stable across surveys.}
\includegraphics[width=0.95\columnwidth]{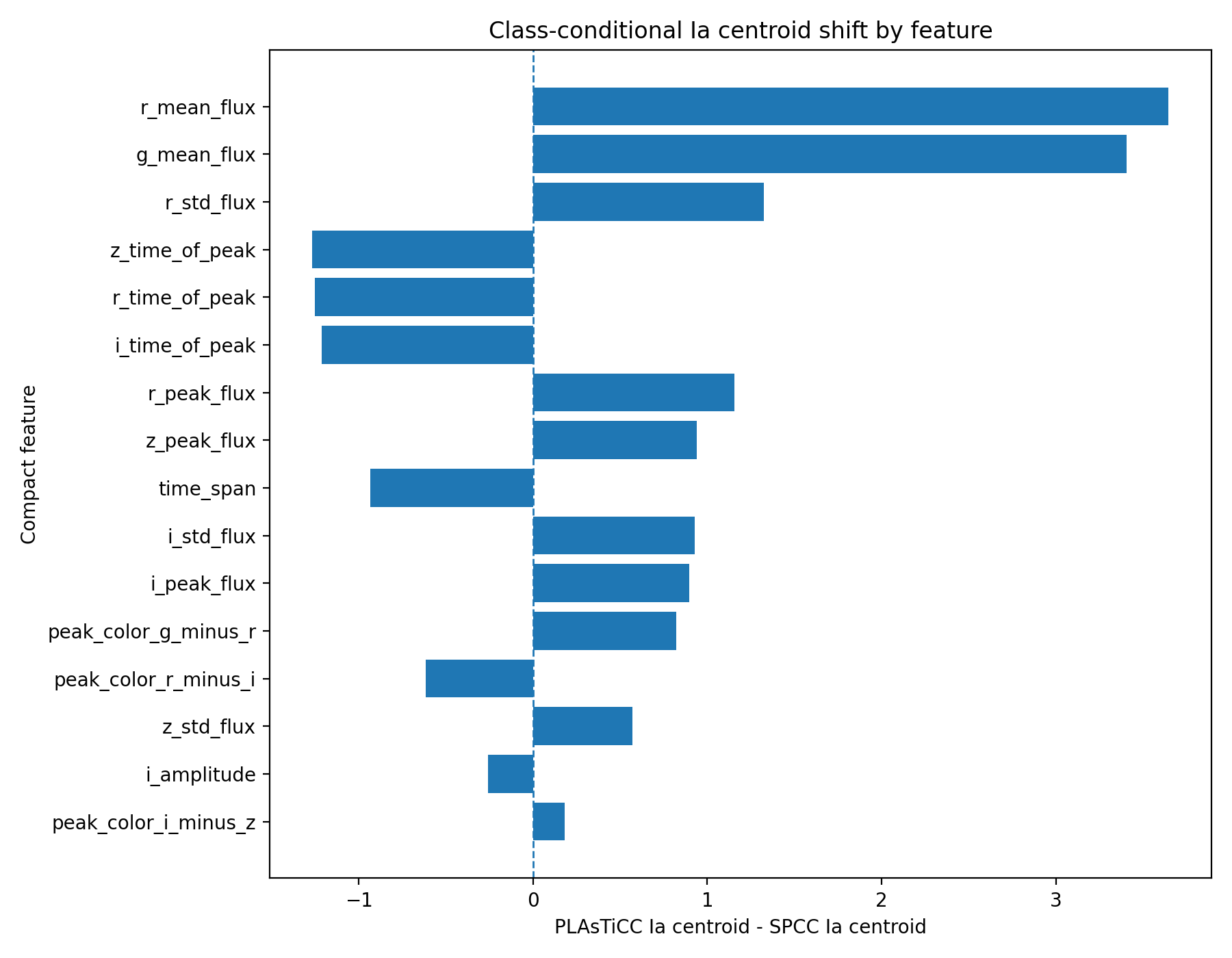}
\label{fig:centroid_shift_barplot}
\end{figure}

The key result, visible in Fig.~\ref{fig:centroid_pca} and quantified in Table~\ref{tab:centroid_analysis}, is that the cross-survey Ia centroid shift is larger than the Ia/non-Ia separation within either survey. The SPCC Ia and PLAsTiCC Ia populations therefore do not occupy the same region of the compact feature space. This demonstrates that the transfer gap is not merely a classifier-boundary problem. Instead, the compact feature representation itself is affected by survey-dependent domain shift.

The feature-wise shifts shown in Fig.~\ref{fig:centroid_shift_barplot} further indicate that this mismatch is not uniform across the compact representation, but is concentrated more strongly in some features than others.

This is a negative result for the strongest form of the portability hypothesis. While the compact representation remains robust within SPCC and under controlled perturbations, it does not transfer directly to PLAsTiCC without substantial feature-space mismatch. The result therefore refines the original hypothesis: physical interpretability is necessary for transparent classification, but it is not sufficient for survey invariance.

This result revises the interpretation of cross-survey generalization. Compact physically interpretable features are robust within a survey and useful for diagnostic analysis, but they are not automatically survey-invariant. Cross-survey use requires feature-space harmonization, domain adaptation, or restriction to a smaller set of invariant compact features.

\section{Discussion}

The results presented in this work provide a more nuanced picture of compact physically interpretable feature representations than the original compact-feature study alone. Within the SPCC domain, the compact 16-feature representation remains consistently strong. Performance is stable across multiple nonlinear classifier families, repeated resampling protocols, and a range of controlled perturbations. Even under reduced observations, moderate flux noise, and shortened temporal coverage, the degradation is generally gradual rather than catastrophic. These results support the conclusion that the compact representation captures a genuine classification signal rather than exploiting a fragile feature configuration.

The cross-survey experiments, however, reveal an important limitation. Direct transfer from SPCC to PLAsTiCC produces a substantially larger performance drop than any of the controlled within-domain perturbations. This result initially suggests that the compact representation may not generalize, but the subsequent diagnostics provide a more specific interpretation. Mixed-domain training partially recovers performance, indicating that the transfer gap is not caused solely by classifier failure. Likewise, feature-definition audits and transient-window harmonization show that differences in feature construction and temporal representation contribute to the observed mismatch.

The class-conditional centroid analysis provides the strongest evidence regarding the source of the transfer limitation. The separation between the SPCC and PLAsTiCC Type~Ia centroids exceeds the Ia/non-Ia separation within either survey. In practical terms, this means that Type~Ia events from the two surveys occupy different regions of the compact feature space before any classifier decision boundary is considered. The transfer problem therefore originates primarily from feature-space alignment rather than from optimization of the classification model.

This finding has broader implications for interpretable machine learning in astronomy. Physical interpretability and survey invariance are related but distinct properties. A feature can correspond to a meaningful physical quantity while still exhibiting survey-dependent numerical behavior due to cadence differences, filter response, noise characteristics, selection effects, or feature-construction choices. Consequently, the existence of an interpretable feature representation does not guarantee that a classifier trained on one survey can be deployed directly on another.

This interpretation is important because SPCC and PLAsTiCC are intended to represent the same underlying astrophysical transient classes. The measured centroid displacement should therefore not be read primarily as evidence that Type~Ia supernovae are physically different in the two datasets. A more plausible interpretation is that the same physical process is mapped into different numerical feature regions by survey-dependent observing and processing choices. These include instrumental effects such as filter transmission, detector response, depth, and noise properties; observational effects such as cadence, season length, and redshift coverage; and procedural effects such as active-window definition, band-coverage requirements, and color-feature construction. In this sense, the domain shift is not merely a statistical displacement of two otherwise equivalent distributions, but may reflect differences in the effective measurement process itself.

The mechanism by which this occurs is straightforward in observational terms. A photometric feature is not a direct measurement of the intrinsic supernova light curve alone; it is a summary of the intrinsic spectral energy distribution after convolution with a survey-specific observing system. For example, the flux measured in a band is effectively determined by the transient spectrum, the filter transmission curve, the detector response, atmospheric throughput, cadence, exposure depth, and the epochs at which the transient is sampled. A color feature such as $C_{g-r}$ or $C_{i-z}$ therefore combines astrophysical temperature evolution with the wavelength response and signal-to-noise properties of the observing system. If two surveys use different effective passbands or sample different rest-frame phases, the same underlying Type~Ia event can acquire different compact color coordinates.

Temporal features are affected in a similar way. The measured time of peak, duration of the active window, and band-dependent peak ordering depend on the cadence and observing season, not only on the physical rise and decline of the supernova. A survey with denser sampling around maximum light is more likely to recover peak-related quantities accurately, while a survey with sparse or uneven cadence may shift the measured peak time or compress the apparent active interval. Thus, differences between DES-like SPCC sampling and LSST-like PLAsTiCC sampling can move the centroid of a class even when the underlying transient physics is unchanged.

This is why the centroid displacement is scientifically meaningful rather than merely a numerical artifact. It shows that the compact features encode both astrophysical information and the survey transfer function that maps that astrophysics into measured photometric summaries. The failure of direct SPCC$\rightarrow$PLAsTiCC transfer therefore indicates that a physically interpretable feature is not automatically an instrument- and procedure-independent observable. For cross-survey use, the relevant question is not only whether a feature has physical meaning, but whether it preserves that meaning under different observing strategies and feature-construction choices.

The negative transfer result should therefore be interpreted as a refinement of the original hypothesis rather than a contradiction of it. The compact representation remains valuable because it enables diagnosis of the transfer failure itself. The centroid analysis, feature-shift ranking, and harmonization experiments provide direct insight into why performance degrades across surveys. Such diagnostic capability is substantially more difficult to obtain from opaque latent representations.

\section{Limitations and Future Work}

Several limitations should be considered when interpreting the results of this study. First, both SPCC and PLAsTiCC are simulated survey datasets. Although they are widely used benchmarks for photometric supernova classification, they do not fully reproduce all observational effects, calibration uncertainties, selection biases, and operational complexities present in real survey data. Consequently, the observed domain shift should be interpreted as evidence within the context of these benchmark surveys rather than as a definitive statement about all future survey combinations.

Second, the analysis is intentionally restricted to a fixed compact feature schema derived from physically interpretable light-curve summaries. The negative transfer result therefore applies to the present feature representation and not necessarily to all compact-feature approaches. Alternative feature definitions, different color constructions, survey-specific calibrations, or revised temporal representations may produce stronger cross-survey alignment.

Third, the study focuses primarily on direct transfer rather than on explicit domain adaptation. While transient-window harmonization and mixed-domain training were explored, no dedicated domain-adaptation framework was applied. Methods such as feature-space alignment, invariant representation learning, transfer learning, or survey-aware calibration may reduce the centroid mismatch identified in the present work. However, the present results also suggest that simple distribution matching may not be sufficient if features with the same name do not have the same effective measurement meaning across surveys. Aligning two statistical distributions can make the feature values appear more similar, but it does not necessarily restore equivalence in the underlying observational process that generated those values.

Finally, the class-conditional centroid analysis demonstrates that the compact feature spaces of SPCC and PLAsTiCC are misaligned, but it does not by itself identify the optimal correction strategy. The analysis is diagnostic rather than prescriptive. Additional work is required to determine whether a smaller invariant feature subset, improved harmonization procedures, or more sophisticated adaptation methods can recover survey portability.

Future work should therefore focus on identifying a smaller subset of survey-stable compact features and on testing feature-space harmonization strategies that explicitly minimize cross-survey centroid mismatch. Such work should distinguish between genuine astrophysical invariance and numerical alignment produced by survey-dependent measurement procedures. If successful, these approaches may preserve the interpretability advantages of compact representations while improving survey portability.

The negative transfer result should therefore be interpreted as evidence against direct portability of the present compact representation, not as evidence against all compact-feature or physically interpretable approaches.

\section{Conclusion}

This work tested whether a compact physically interpretable feature representation for Type~Ia supernova classification remains stable beyond the SPCC domain in which it was originally developed. Within SPCC, the compact 16-feature XGBoost model retains strong performance, with F1 $\approx 0.844$ and PR-AUC $\approx 0.928$, and its behaviour remains stable across alternate classifiers, repeated resampling, and several controlled feature-space perturbations.

The strongest limitation appears under external survey transfer. Direct SPCC$\rightarrow$PLAsTiCC evaluation produces a substantially larger performance drop than the controlled SPCC-like perturbations. Mixed-domain training and transient-window harmonization partially reduce this gap, showing that the compact representation is not entirely uninformative outside SPCC. However, class-conditional centroid analysis demonstrates that SPCC and PLAsTiCC Type~Ia populations do not occupy the same region of the standardized compact feature space. The cross-survey Type~Ia centroid shift is larger than the Ia/non-Ia separation within either survey.

The strongest form of the portability hypothesis is therefore not supported. Compact physically interpretable features are useful for within-survey classification and for diagnosing why transfer fails, but they are not automatically survey-invariant. The main conclusion is that physical interpretability is necessary for transparent classification, but it is not sufficient for direct cross-survey deployment.

A defensible path forward is not to treat the current compact model as a universal classifier, but to use it as a diagnostic framework for identifying survey-stable features. Future work should therefore test invariant feature subsets, feature-space harmonization, and explicit domain-adaptation methods before claiming cross-survey portability, while distinguishing genuine astrophysical invariance from numerical alignment produced by survey-dependent measurement procedures.

\appendix

\section{Summary of Transferability Diagnostics}
\label{app:diagnostics}

\begin{table*}[t]
\centering
\caption{Summary of transferability and robustness diagnostics used in this work.}
\label{tab:transferability_design}
\begin{tabular}{p{0.20\textwidth}p{0.36\textwidth}p{0.34\textwidth}}
\hline
Diagnostic & Evidence used & Interpretation \\
\hline
Classifier-family comparison & XGBoost, Random Forest, SVM, and Logistic Regression were evaluated on the same compact feature set (Table~\ref{tab:model_compare}). & The compact signal is not exclusive to one classifier, although XGBoost provides the strongest PR-AUC. \\
Resampling stability & Repeated $k$-fold and random-split experiments produced small F1 variation around the compact baseline (Table~\ref{tab:resampling_stability}). & The compact baseline is not strongly dependent on one favourable train/test split. \\
Controlled SPCC perturbations & Noise, reduced observations, missing-band proxies, flux scaling, and shortened temporal span were tested (Table~\ref{tab:perturbation_robustness}). & The feature set degrades gradually under internal perturbations, with missing red-band information and shortened coverage producing the largest drops. \\
Domain-swap transfer & SPCC-trained models were evaluated on shifted SPCC variants and PLAsTiCC compact features (Table~\ref{tab:domain_swap}). & PLAsTiCC transfer is substantially harder than controlled SPCC-like perturbations. \\
Mixed-domain training & Models were trained on SPCC alone and on mixtures including shifted SPCC variants and PLAsTiCC-like examples. & Adding PLAsTiCC-like examples improves average cross-domain performance, showing that part of the gap is train/deployment mismatch rather than total feature failure. \\
Feature-definition audit & The SPCC and PLAsTiCC compact-feature construction pipelines were compared feature by feature. & Several features had the same formal definition but different effective meaning because the surveys initially represented different temporal domains. \\
Normalization and alignment trials & Event-level light-curve normalization, temporal-offset variants, and PCA-alignment checks were tested. & Simple amplitude normalization, relative-time reparameterization, or linear feature-space rotation did not remove the transfer gap. \\
Transient-window harmonization & PLAsTiCC compact features were rebuilt using transient-centered windows and a window sweep over multiple half-widths. & Matching the PLAsTiCC extraction window to the SPCC event-duration scale improved transfer, indicating that temporal-domain mismatch is a real contributor. \\
Class-conditional centroid test & SPCC and PLAsTiCC Ia/non-Ia centroids were compared in standardized compact-feature space (Fig.~\ref{fig:centroid_pca}, Fig.~\ref{fig:centroid_shift_barplot}, Table~\ref{tab:centroid_analysis}). & The remaining transfer limitation is a feature-space alignment problem, not only a classifier-boundary problem. \\
\hline
\end{tabular}
\end{table*}

\section*{Acknowledgements}
The author acknowledges the public availability of the SPCC dataset and the open-source scientific Python ecosystem that made this analysis possible.

\section*{Funding}
This research received no specific grant from any funding agency in the public, commercial, or not-for-profit sectors.

\section*{Author Contributions}
Anurag Garg conceived the study, designed the methodology, performed the analysis, interpreted the results, and wrote the manuscript.

\section*{Data Availability}
This study uses the Supernova Photometric Classification Challenge (SPCC) dataset. The processed feature tables and analysis scripts used in this work are available in the public repository \url{https://github.com/mranuraggarg/supernovae_classification}. The manuscript-specific materials associated with the present analysis are archived and cited as \cite{GargRepo2026}.

\section*{Code Availability}
The code used for feature extraction, model training, and feature ablation analysis is available at \url{https://github.com/mranuraggarg/supernovae_classification}. The workflow corresponding to the present manuscript is archived and cited as \cite{GargRepo2026}.

\section*{Conflict of Interest}
The author declares no conflict of interest.

\bibliographystyle{apj}
\bibliography{references}

@article{Kessler2010,
  author = {Kessler, R. et al.},
  title = {Results from the Supernova Photometric Classification Challenge},
  journal = {PASP},
  year = {2010},
  volume = {122},
  pages = {1415},
  doi = {10.1086/657607}
}

@article{Lochner2016,
  author = {Lochner, M. and McEwen, J. D. and Peiris, H. V. and Lahav, O. and Winter, M. K.},
  title = {Photometric Supernova Classification With Machine Learning},
  journal = {ApJS},
  year = {2016},
  volume = {225},
  pages = {31},
  doi = {10.3847/0067-0049/225/2/31}
}

@article{Karpenka2013,
  author = {Karpenka, N. V. and Feroz, F. and Hobson, M. P.},
  title = {Automated photometric classification of supernovae},
  journal = {MNRAS},
  year = {2013},
  volume = {429},
  pages = {1278},
  doi = {10.1093/mnras/sts412}
}

@article{Charnock2017,
  author = {Charnock, T. and Moss, A.},
  title = {Deep recurrent neural networks for supernova classification},
  journal = {ApJL},
  year = {2017},
  volume = {837},
  pages = {L28},
  doi = {10.3847/2041-8213/aa5bf7}
}

@article{Moller2019,
  author = {Möller, A. and de Boissière, T.},
  title = {SuperNNova: an open-source framework for Bayesian, Neural Network based supernova classification},
  journal = {MNRAS},
  year = {2019},
  volume = {491},
  pages = {4277},
  doi = {10.1093/mnras/stz3062}
}

@article{Pasquet2019,
  author = {Pasquet, J. and Pasquet, J.},
  title = {Deep learning for supernova classification},
  journal = {A\&A},
  year = {2019},
  volume = {611},
  pages = {A97},
  doi = {10.1051/0004-6361/201731992}
}

@article{Garg2026,
  author = {Garg, Anurag},
  title = {Compact and Physically Interpretable Feature Models for Photometric Type Ia Supernova Classification},
  journal = {Journal of Astrophysics and Astronomy},
  year = {2026},
  note = {accepted for publication in Journal of Astrophysics and Astronomy; arXiv:2603.14500},
  doi = {10.48550/arXiv.2603.14500}
}

@misc{GargRepo2026,
  author = {Garg, Anurag},
  title = {supernovae\_classification: Compact feature models for photometric supernova classification},
  year = {2026},
  howpublished = {\url{https://github.com/mranuraggarg/supernovae_classification}},
  note = {GitHub repository}
}

@article{Ivezic2019,
  author = {Ivezi\'c, \v{Z}. and others},
  title = {LSST: From Science Drivers to Reference Design and Anticipated Data Products},
  journal = {ARA\&A},
  volume = {57},
  pages = {417-487},
  year = {2019},
  doi = {10.1146/annurev-astro-091918-104453}
}

@book{Hastie2009,
  author    = {Hastie, Trevor and Tibshirani, Robert and Friedman, Jerome},
  title     = {The Elements of Statistical Learning: Data Mining, Inference, and Prediction},
  edition   = {2},
  year      = {2009},
  publisher = {Springer},
  address   = {New York},
  doi       = {10.1007/978-0-387-84858-7}
}

@article{Garg2025,
  author  = {Garg, Anurag},
  title   = {Optimizing supernova classification with interpretable machine learning models},
  journal = {Journal of Astrophysics and Astronomy},
  year    = {2025},
  volume  = {46},
  pages   = {86},
  doi     = {10.1007/s12036-025-10113-4}
}

\end{document}